# Effective Mg Incorporation in CdMgO Alloy on Quartz Substrate Grown by Plasma-assisted MBE

A. Adhikari*,[1], A. Wierzbicka[1], A. Lysak[1], P. Sybilski[1], B. S. Witkowski[1], and E. Przezdziecka[1]

[1]*Institute of Physics, Polish Academy of Sciences, Al. Lotnikow 32/46, Warsaw, Poland, 02-668*

Email: adhikari@ifpan.edu.pl

## Abstract

The development of CdMgO ternary alloy with a single cubic phase is challenging but meaningful work for technological advancement. In this work, we have grown a series of $Cd_{1-x}Mg_xO$ ternary random alloys with various Mg concentrations (x = 0, 30, 32, 45, and 55%) on quartz substrate by plasma-assisted molecular beam epitaxy (PA-MBE) technique. The structural investigations of alloys were performed using the X-ray diffraction (XRD) technique. The decreases in average crystallite size and lattice parameters were observed with an increase in Mg content in the alloys. XRD analysis confirms a single cubic phase is obtained for alloy compositions. The elemental and morphological studies were carried out using energy dispersive x-ray (EDX) spectroscopy and atomic force microscope (AFM) technique, respectively. The optical investigation was carried out using UV-Vis spectroscopy. The optical bandgaps were estimated using the Tauc relation and it was varied from 2.34 eV to 3.47 eV by varying the Mg content from zero to 55% in the alloys. The Urbach energy increases from 112 meV to 350 meV which suggests a more disordered localized state with an increase in Mg incorporation in the alloys.

# 1. Introduction

The oxide semiconductors have gained more attention among the other semiconductor materials because of their potential applications in optoelectronic industries like flat panel displays (FPDs) [1], thin-film transistors (TFTs) [2], photodetectors [3,4], LEDs [5], sensors [6,7], etc. Metal oxide semiconductors that have a large bandgap energy value, have a better tolerance against high voltage. Apart from that, these are relatively stable in an atmospheric environment which can make use of the fabrication process easy [8]. Among the oxide semiconductors, group-II oxides receive increasing interest as they share similar features with group-III nitrides which could be very promising for semiconductor industries. From the early 90's research on wurtzite ZnO as an alternative to the GaN system has started and still is going on because of its wide direct band gap of 3.37 eV and large exciton binding energy of 60 meV which is higher than in GaN (binding energy 25 meV) [9]. The rise in research not only about the basic properties of ZnO but also on the related group II-VI semiconductor oxides like MgO and CdO which can be combined to form heteroepitaxial device structure. The large bandgap tunability, stable structure, and ease of fabrication process make the ZnO-CdO-MgO system a possible alternative to nitrides [10].

Among these metal-semiconductor oxides, CdO is considered to be a promising material for photovoltaic applications due to its unique features like stable cubic rock-salt (RS) structure, high transparency, high exciton binding energy of 75 meV, a high dielectric constant of 21.9, and high electron mobility of 180 $cm^2V^{-1}s^{-1}$ [11]. However, a relatively small direct bandgap of 2.3 eV along with two indirect bandgaps of 1.2 eV and 0.8 eV [12] at room temperature is a key problem that restricts its uses in a shorter wavelength regime. On the other hand, MgO is a direct bandgap (7.5 eV) material with a high exciton binding energy of 80 meV [13] that crystallizes in the cubic rocksalt structure. Thus, MgO can be combined with CdO by the meaning of alloying to form various heterostructures such as ternary alloys, CdO/MgO superlattice structures (SLs), etc. which can be useful for a variety of optoelectronic devices over a wide spectral range. Previously Chen *et al.* [14] studied the change in the electronic structure of CdMgO alloys on glass substrates prepared by radio frequency magnetron sputtering technique. Guia *et al.* studied Mg-rich to Cd-rich CdMgO alloys on a sapphire substrate prepared by the metal-organic chemical vapor deposition (MOCVD) technique and reported a phase separation at Cd concentration above 27% [15,16]. Przezdziecka *et al.* studied quasi alloys short-period {CdO/MgO} SLs grown on sapphire substrate by plasma-assisted molecular beam epitaxy (PA-MBE) technique and found out that the bandgap of SLs can be varied from 2.6 eV to 6 eV by changing the CdO sublayer thick-ness at a constant MgO layer thickness [17,18]. Recently, Adhikari *et al.* reported $Cd_xMg_{1-x}O$ ternary alloys on m- and *c*-plane sapphire substrate grown by PA-MBE technique. They have found that the direct bandgap can be varied from 2.42 eV to 5.5 eV by changing Mg content from 0 to 96% [19].

For the design of optoelectronic devices, optical coatings, and the manufacture of solar cells, a limited number of suitable materials are available. Few theoretical articles have been reported in recent years that indicate the possible tunability of the energy band gap of CdMgO and stability of the cubic rock-salt structure across a full range of magnesium which can prove useful from a practical point of view [20]. The development of new oxide-based devices requires improving our knowledge about ternary alloys like CdMgO. Previously, there have been some studies for quasi-ternary alloys based on ZnO/MgO [21,22] and ZnO/CdO [23,24] structures where composition inhomogeneity can be overcome by growing superlattice structures. However, there is still limited work being carried out on CdO-MgO alloys [25–27]. In this paper, we have exploited CdMgO layers deposited on quartz substrate by plasma-assisted MBE technique. In particular, in the MBE method obtained layers are hydrogen-free. In distinguishing from other methods e.g.: MOCVD where metalorganic precursors are the source of the hydrogen and carbon contamination in the samples, the number of contaminations in the MBE method is low. Both the substrates and the growth method influence the crystallographic and optoelectrical parameters of the semiconductors [28–30]. In particular, crystalline substrates can impose a preferential layer orientation, as it was observed for CdMgO layers on differently oriented sapphire substrates [19]. It is, therefore, reasonable to characterize CdMgO samples obtained by different methods grown on different substrates including amorphous substrates. We have chosen the quartz substrate as a cheaper alternative to the sapphire substrate for our study. Moreover, the amorphous substrate does not dictate preferential orientation in CdMgO films by the orientation of the substrate. The forced orientation on amorphous substrates is related only to the thermodynamics of growth. Simultaneously, the use of a transparent quartz substrate allows for a relatively easy study of the energy gap of the layers by the optical transmission method. The Mg content in the alloys was varied by changing the growth condition e.g.: Mg and Cd fluxes ratio. Structural, morphological, and optical investigations of these samples were carried out in order to deeper characterize CdMgO ternary material.

## 2. Experimental Methods

A Riber Compact 21B MBE system equipped with Mg and Cd effusion cells, and radiofrequency oxygen plasma cells were used to grow CdMgO random alloys on a quartz substrate. As the sources, Mg (6N purity, from PUREMAT technologies) and Cd (6N purity, from JX Nippon Mining & Metal Corporation) ingots were used. The substrates were cleaned with acetone and next with isopropanol for 20 minutes followed by deionized water in order to remove organic contaminations and next dried with pure 5N $N_2$ gas. During the growth process, the Cd and Mg fluxes were controlled by varying the effusion cell's temperatures. The temperature of the Cd effusion cell was fixed at 380°C whereas, the Mg effusion cell temperature was varied from $T_{Mg}$= 500°C, 510°C, 520°C, and 530°C for samples S2 to S4 respectively. The growth process was carried out at 360oC measured by thermocouple with a constant oxygen flow of 2.5 ml/min at a fixed 400 W radio frequency power of the oxygen plasma.

The structural investigations of these ternary alloys were performed using a PANalytical X'Pert Pro MRD, X-ray diffraction technique. Cu anode used for CuKα1 radiation of 1.5406Å wavelength. The diffractometer is equipped with a Ge analyzer and a 2-bounce Ge (200) hybrid monochromator. Elemental compositions in the alloys were investigated using a Hitachi SU-70 scanning electron microscope equipped with Thermo Scientific energy-dispersive X-ray (EDX) spectrometer. The excitation energy was set at 6 keV, which allowed for effective excitation and measurement of the composition of magnesium and cadmium, and at the same time minimized the generation of a signal from the substrate that could disturb the measurement. The change in growth parameter (i.e., Mg effusion cell temperature) results in, different Mg content $Cd_{1-x}Mg_xO$ ternary alloys. The surface morphology was studied using Atomic force microscopy (AFM) in tapping mode (Bruker Dimension Icon model). The optical investigations of these alloys were performed at room temperature using a Carry 5000 UV-Vis/NIR spectrophotometer equipped with a PbS detector from Agilent Technologies in a 200-600 nm wavelength range.

## 3. Results and Discussions

*3.1. Structural property*

Figure 1(a) shows the structural properties of CdMgO alloys on quartz substrate using X-ray diffraction (XRD) 2θ scan over a range from 20º to 100º. XRD results confirmed that all the samples were cubic rocksalt (RS) structures without any additional impurity phases. For pure CdO (S1), the 111 and 200 diffraction peaks are identified from the diffraction angle at about 33º and 39º respectively-indicates the polycrystalline character of the structure. These identified reflections are to originate from the cubic RS structure of CdO and are in good consistency with X-ray diffraction JCPDF card no. 652908. Furthermore, there are no additional peaks observed co-related to the quartz substrate in XRD analysis. XRD peak positions of pure CdO and MgO with the cubic structure are marked with a dotted line in Fig. 1(a). A systematic shift of (111) and (200) diffraction peaks towards a higher diffraction angle have been observed with an increase in Mg content in the alloy (Fig. 1(b)). In case of layers obtained by pulsed laser deposition (PLD) technique on quartz, the films grown at low-temperature show preferred orientation along (111) direction, while the films grown at high temperature have preferred orientation along (200) direction [31]. The two orientations 111 and 200 were reported in a temperature range of about 400ºC. Our analysis indicated that preferential crystallographic orientation was also correlated with Mg concentration in the alloy. The areal comparison of 111 and 200 diffraction peaks is shown in Fig. 1(c). Due to the increase of the Mg content, the contribution from the 111 orientation increases compared to 200 orientation which reveals that the preferential orientation is strongly influenced by the Mg content in the alloys. Previously, Adhikari *et al.* [19] reported the presence of double XRD peaks in $Cd_xMg_{1-x}O$ film on m- and c-plane sapphire grown by plasma-assisted MBE technique. It was evidence of the existence of regions with two different concentrations in the CdMgO ternary layers. Their study suggested that both growth conditions and the orientation of substrates play an important role in the homogeneity of the CdMgO layer. However, in this study, no such behavior has been observed. It was observed that the FWHM of XRD CdMgO peaks on quartz substrates are much higher than it was reported for sapphire substrates. The observed broadening of the diffraction peak with an increase in Mg content in the alloy, clearly indicates that the incorporation of Mg ion into the CdO lattice can be not fully homogenous. Moreover, the noisy diffraction peak suggests the not-so-good crystallinity of the prepared alloy on the quartz substrate compared to previously studied samples on sapphire substrates [19].

As CdO and MgO both crystallize in the cubic rocksalt structure, in CdMgO alloys each oxygen atom is surrounded by Cd or Mg atoms and is octahedrally coordinated to each other at an equivalent position. The lattice parameter can be determined from Bragg's equation

$$2d\ sin\theta = n\ \lambda \qquad (1)$$

where $d$ is interplanar spacing, n is the order of diffraction (usually $n=1$), $\lambda$ is the X-ray wavelength used, and $\theta$ is the Bragg angle. For cubic structure, the interplanar distance d is related to lattice constant a with relation,

$$a = d * (h^2 + k^2 + l^2)^{\frac{1}{2}} \qquad (2)$$

where $h, k$, and $l$ are Miller indices of diffraction planes. For (111) diffraction plane, the lattice constant a can be calculated using the relation, $a=(\sqrt{3}\lambda)/2sin\theta$. The calculated lattice parameters of CdMgO alloys are listed in Table 1. The shifting of 111 diffraction peak towards a higher diffraction angle results from decreasing lattice parameters with increasing Mg content in the alloys. It can be expected due to the substitution of a smaller atomic radius of Mg than Cd, similar to the case of AlGaN [32]. The calculated lattice parameters show a nearly linear behavior in the analyzed Mg content. Previously, reported values of the bowing parameters in the literature were found to be, b = -0.449 nm for MgCdO/$c$- Al$_2$O$_3$ [16], -0.89 nm for CdMgO/$m$- Al$_2$O$_3$, and -1.05 nm for CdMgO/$c$- Al$_2$O$_3$ [19]. In order to determine the microstructural contribution such as average crystallite size and microstrain, here we have adopted a size-strain plot [33,34]. The crystallite size and microstrain in the alloys were estimated by fitting the 111 diffraction peak using the Voigt function. The observed broadening in the diffraction peak is due to the combined effect of crystallite size as well as the lattice microstrain in the alloys. The Gaussian component of the FWHM ($\beta_G$) analysis is used for determining strain whereas the Lorentzian component of FWHM ($\beta_L$) is used for determining average crystallite size or grain sizes ($\beta_{hkl}=\beta_L+\beta_G$). The average crystallite size is estimated using the Debye-Scherrer relation,

$$\tau = \frac{k\lambda}{\beta_L\ cos\theta} \qquad (3)$$

where $\tau$ is the average crystallite size, $k$ is a constant of value 0.94, and $\beta_L$ is the Lorentzian component FWHM of diffraction peak. As shown in Fig. 2 the grain size decreases with an increase in Mg content in the alloy. A similar trend with smaller grain size varying from 13 nm to 4 nm with an increase in Mg content 0% to 28% in CdMgO alloy on a glass slide was reported by Chen *et al.* [14]. However, a higher grain size was observed for CdMgO alloys on both m- and c-plane sapphire substrates prepared by the PA-MBE technique [19]. The other microstructural contribution such as lattice microstrain arises due to various factors such as the constituent atoms (Cd and Mg), substrate material, and growth temperature in the alloys [16]. The microstrain ($\varepsilon$) in the alloy can be estimated from strain-induced broadening using the relation,

$$\varepsilon = \frac{\beta_G}{4\tan\theta} \quad (4)$$

The Lorentzian and Gaussian components of FWHM were estimated along with the instrumental errors equal to $\delta\beta_L \pm 0.0014°$, and $\delta\beta_G \pm 0.0035°$ respectively. Table 1 illustrates the average grain size and the microstrain in the alloys along 111 diffraction peak directions. It is observed that micro strain increases with an increase in Mg content in the alloys as shown in Fig. 2. The obtained values of microstrain are comparable with previously reported CdMgO alloys on sapphire substrates [19] grown by PA-MBE technique, and Mg-doped CdO films obtained by spray pyrolysis technique [35]. Previously, a comparatively higher value of microstrain is reported at low growth temperature (Cd-rich) and it decreases with an increase in nominal Cd concentration in CdMgO alloys on a c-sapphire substrate prepared by the MOCVD technique [16]. However, in all the cases an effective lattice shrinkage and an increase of the lattice microstrain with further increase in Mg incorporation suggest better crystalline quality alloy at Cd-rich CdMgO conditions. It is worth noting that, the dominant contribution of the diffraction peak broadening comes from grain size in the case of Cd-rich conditions whereas with increases in Mg content the major contribution comes from microstrain in the alloys. The obtained microstrain can lead to forming localized states and is expected to affect both electrical and optical properties in CdMgO alloys.

*3.2. Elemental and Morphological study*

The elemental distribution, as well as Cd and Mg composition in the alloys, was studied using the EDX technique. The Mg content in the alloy was varied by changing the growth temperature of the Mg effusion cells. By the EDX method, the total amount of elements in samples was measured. X-ray examinations did not reveal the presence of metallic inclusions in the tested samples. In X-ray diffraction, we do not see the amorphous signal as well as any signals correlated with the presence of other oxides. The atomic percentage values of Cd and Mg contents are listed in Table 2. The Mg content in the alloy can be increased by increasing the Mg effusion cell temperature. The elemental distributions of Cd and Mg atoms in the alloys are shown in Fig. 3. The EDX spectra were collected from over the samples as shown in Fig. 3. The EDX mapping confirms a uniform homogeneous distribution (in the presented scale) of Cd and Mg elements in the corresponding alloys. All of the Cd and Mg content measurements were performed within an error margin of ±2%.

The analysis of the nanotexture of the front transparent conductive oxide (TCO) electrode and/or back reflector used in solar cells is very important. For such photovoltaic applications, the average roughness parameter, $R_a \ll \lambda$ ($\lambda$ is the light wavelength in a vacuum) is known as a criterion for the light scattering and trapping to achieve high efficiency of the devices [36]. Thus, the surface morphology of CdMgO ternary alloys with various Mg content was studied using AFM. Figure 4 shows the surface scan over a 5x5 µm$^2$ area of ternary alloys. The average roughness parameters ($R_a$) have been calculated using Nanoscope software and are listed in Table 2 with a variation of Cd and Mg content in the alloy. The smallest value of the $R_a$ is obtained for samples with the highest concentration of Mg. The decrease in the average roughness parameter ($R_a$) is implying that a smoothening surface morphology by the change in Mg content in the alloys. The obtained result is consistent with the previously studied CdMgO alloy on the sapphire substrate. It was presented that the variation of the average roughness parameter value for the whole Mg concentration region of CdMgO alloys is within the 2-10 nm range [19]. It was found that in case of CdMgAlO obtained by PLD roughness parameters do not change significantly with growth temperature [31]. However, Karakaya *et al.* reported a smoother surface for 15% and 30% Mg concentration in CdO:Mg films obtained by spray pyrolysis on glass substrates [37]. A better surface uniformity and fewer grain cavities were reported for 4% Mg-doped CdO film prepared by spray pyrolysis on a glass substrate [38]. The variation of the roughness parameter with Mg content on different substrates is shown in Fig. 5(a). In our case, the roughness is strongly correlated with grain sizes. A higher grain size indicates a higher value of $R_a$ (as shown in Fig. 5(b)). A similar correlation between Ra and grain sizes was previously noted in case of CdO doped with Cu [39], CdO doped with Ce [40], and Ni-doped CdO [41].

*3.3. UV-Vis spectroscopy*

The optical transmission spectra of CdMgO alloys on quartz substrate are shown in Fig. 6(a). All the prepared alloys have high transparency (more than about 90%), which suggests that they can be used in many applications as transparent conductive oxide (TCO). The absorption coefficient (α) of the sample is determined from the transmittance spectra using relation [42], $\alpha = -(1/d) \ln (\%T)$, and the corresponding bandgap has been calculated using Tauc relation, $\alpha h\nu = A(h\nu - E_{g,o})^m$, where d is the sample thickness (~300 nm), $h\nu$ is the photon energy in eV, $E_{g,o}$ is the optical bandgap, and A and m are constants. The m value depends upon the type of electronic transition in the semiconductors. The value of m is equal to ½ or 2 for direct or indirect transition in the semiconductor, respectively. The optical direct and indirect bandgap of the films were calculated by extrapolating the linear portion of $\alpha^2$ and $\alpha^{1/2}$ with the energy axis as shown in Fig. 6(b) and Fig. 6(c). This procedure was usually adopted for calculations of energy gaps in semiconducting materials [43–46]. The direct bandgap of pure CdO (S1) is found to be 2.34 eV. It is observed that the absorption edge shifts toward higher energy with an increase in Mg contentment in the alloy. The direct bandgap changes from 2.34 eV to 3.47 eV with a change in Mg contentment from 0% to 55% in the alloys. MgO has a direct energy gap, hence, as the content of Mg in CdMgO alloys increases, the input from the direct gap becomes dominant and it is difficult to observe the indirect gap and we do not see it for samples with higher Mg content alloys. The indirect bandgap of samples with higher Cd content i.e., S1 (CdO), S2 ($Cd_{0.7}Mg_{0.3}O$), and S3 ($Cd_{0.68}Mg_{0.32}O$) are estimated from Tauc relation and shown in Fig. 6(c). For pure CdO, the indirect bandgap was found to be 1.99 eV which is in good agreement with the experimental bandgap values obtained in the literature [47-48]. However, a lower value of 1.2 and 0.8 eV indirect bandgaps were reported in the literature [12] obtained from the theoretical calculation. The up-shifting of both direct and indirect bandgap with an increase in Mg content depends upon structural as well as electronic contribution in the alloys [14,49]. The widening of the bandgap can be explained using the following interpretations: the top of the valence band in both CdO and MgO is dominated by the O-*2p* state. Furthermore, Cd-*5s* and Mg-*3s* states are dominant at the bottom of the conduction band in the case of CdO and MgO respectively (Fig. 7). Since, the Mg-*3s* state has higher energy compared to the Cd-*5s* state, with further incorporation of Mg into CdO lattice leads the conduction band edge shifts upward with reference to the Fermi level, which results in an increase in bandgap in the alloys [50,51].

The direct bandgap energy of CdMgO can be depicted as a function of Mg content in the alloys which deviates from Vegard's law (Fig. 8(a)). This modification can be expressed using Modified Vegard's law [52] using the relation:

$$(E_g)_{Cd_{(1-x)}Mg_xO} = (1-x)(E_g)_{CdO} + x(E_g)_{MgO} - bx(1-x) \quad (5)$$

where $(E_g)_{CdO}$ and $(E_g)_{MgO}$ are bandgaps of CdO (2.3 eV) and MgO (6 eV) respectively, whereas b is the bandgap bowing parameter. The best-fit results in bandgap bowing parameter of (2.94±0.26) eV as shown in Fig. 8(a). The obtained bowing parameter is comparable with the previously reported bowing constant of 2.17 eV and 3.1 eV for CdMgO alloys on the glass substrate and sapphire substrate respectively [14,25]. The bowing occurs due to volume deformation, charge exchange, and structural relaxation effects in the alloys [53]. However, in our case of cubic CdMgO alloys, it is expected a dominant structural relaxation contribution, where there is a change in bond length as well as bond angle due to the incorporation of foreign Mg atom into the host CdO lattice.

It is well-known fact that semiconductors tend to absorb light at photon energy below the bandgap which strongly depends upon the localized exciton state and energetic disorder. The absorption coefficient varies exponentially with photon energy, below the absorption edge. It can be described with relation, $\alpha = \alpha_0 \exp(h\upsilon / E_U)$, where $\alpha_0$ is the absorption coefficient, $h\upsilon$ is the photon energy, and $E_U$ is the Urbach energy which is related to the defects in the semiconductor material [46,54]. The Urbach energy is determined from the inverse of the slope of the linear portion of ln α with the energy axis. Variation of optical bandgap (direct and indirect bandgap) and Urbach energy with change in Mg content in the alloys are shown in Fig. 8. The error bar is determined by taking the uncertainty in extracting the linear portion of the absorption curve and measurement of Mg content in the alloys. A relatively low value of Urbach energy of 112 meV is found for pure CdO (S1). However, it increases with an increase in the incorporation of Mg into the CdO lattice, which suggests more disorder (Fig. 8 (b)). These disorders can be expected from various localized states which arise with an increase in Mg content. Furthermore, the variation of Urbach energy can be well correlated with structural analysis as the XRD study suggests an increase of microstrain which leads to form various localized states with an increase in Mg content in the alloys. Huso *et al.* [55] reported structural defect and alloy inhomogeneity with an increase in Mg concentration in ZnMgO alloy grown by DC-magnetron sputtering on $CaF_2$ substrate. The corresponding Urbach energy was varied from 80 meV to 250 meV with a change in Mg content 0 to 65% in ZnMgO alloys. However, in this work, it is worth noting that, Mg substitution at the Cd-site introduces the difference between ionic radii of $Mg^{2+}$ and $Cd^{2+}$ which may cause potential fluctuations that may introduce onsite electronic disorder as well as a structural disorder in the alloy [56]. These disorder states appear near the valence band and/or conduction band in the form of Urbach energy ($E_U$) which could scale with higher Mg substitution. From obtained values of $E_U$ (Fig. 8 (b)), it is possible to extract the so-called steepness parameter ($\sigma$) which characterizes the steepness or width of the straight line near the absorption edge. The steepness parameter can be determined from Eq, $E_U = K_B T/\sigma$, where $K_B$ is the Boltzmann constant. Furthermore, the steepness parameter also determines the strength of electron-phonon interaction ($E_{e-p}$) and both are related to each other via the following relationship $E_{e-p}=2/3\ \sigma$ [57,58]. The obtained values of the steepness parameter decrease from Mg concentration and its values are in the range of 0.23 to 0.07 meV. The obtained values are lower than that reported for undoped ZnO at room temperature [59] and higher than values reported for pure CdS obtained by different methods [60]. The electron-phonon coupling depends on the size, structural quality, and concentration of dopants [61].

## 4. Conclusions

A series of ternary alloys of CdMgO with different Mg content was grown on amorphous quartz substrates by the MBE method by controlling the temperature of the Mg cell. All samples of CdMgO alloys reveal the rocksalt cubic structure. The presented XRD results show a systematic shift of the diffraction peaks with increasing Mg content, indicating a decrease in the lattice parameter in CdMgO layers. A reduction in the mean grain size is also observed with an increase in Mg content. It is shown that despite the use of amorphous substrates the layers grow in a preferential orientation, which depends on the Mg content in the alloys. From the EDX maps analysis, a uniform distribution of Cd and Mg atoms was found in all samples. The direct and indirect bandgaps were determined using UV-Vis spectroscopy. The direct bandgap varied from 2.34 eV to 3.47 eV with a change in Mg concentration from 0% to 55%, whereas the indirect bandgap increased from 1.99 eV to 2.3 eV with a change in Mg content from 0% to 32% in the alloys. The obtained values of bandgaps were fitted with modified Vegard's law and the bowing parameter was estimated to be (2.94±0.26) eV. The Urbach energy and microstrain increase with an increase in Mg content in the alloys, which suggests a more disordered film for Mg-rich CdMgO films.

**Acknowledgment**

This was funded in whole by the National Science Center, Poland Grant No. 2021/41/N/ST5/00812, and 2021/41/B/ST5/00216. For the purpose of Open Access, the author has applied a CC-BY public copyright license to any Author Accepted Manuscript (AAM) version arising from this submission. We thank Prof. Adrian Kozanecki, for the scientific discussion and for carefully reading our manuscript.

**References**

[1] J.F. Wager, B. Yeh, R.L. Hoffman, and D.A. Keszler, *Curr. Opin. Solid State Mater. Sci.* **18**, 53 (2014). DOI: 10.1016/J.COSSMS.2013.07.002.

[2] H. Yabuta, M. Sano, K. Abe, T. Aiba, T. Den, H. Kumomi, K. Nomura, T. Kamiya, and H. Hosono, *Appl. Phys. Lett.* **89**, 112123 (2006). DOI:10.1063/1.2353811.

[3] H. Zhu, C.X. Shan, L.K. Wang, J. Zheng, J.Y. Zhang, B. Yao, and D.Z. Shen, *J. Phys. Chem. C* **114**, 7169 (2010). DOI:10.1021/JP101083N.

[4] W. Ouyang, F. Teng, J.-H. He, X. Fang, W.X. Ouyang, X.S. Fang, F. Teng, and J.-H. He, *Adv. Funct. Mater.* **29**, 1807672 (2019). DOI:10.1002/ADFM.201807672.

[5] T. Arai, N. Morosawa, K. Tokunaga, Y. Terai, E. Fukumoto, T. Fujimori, T. Nakayama, T. Yamaguchi, and T. Sasaoka, *SID Symp. Dig. Tech. Pap.* **41**, 1033 (2010). DOI:10.1889/1.3499825.

[6] J. Zhang, Z. Qin, D. Zeng, and C. Xie, *Phys. Chem. Chem. Phys.* **19**, 6313 (2017). DOI:10.1039/C6CP07799D.

[7] G.F. Fine, L.M. Cavanagh, A. Afonja, and R. Binions, *Sensors (Basel)*. **10**, 5469 (2010). DOI :10.3390/S100605469.

[8] M. Kimura, *Jpn. J. Appl. Phys.* **58**, 090503 (2019). DOI :10.7567/1347-4065/ab1868.

[9] Ü. Özgür, Y.I. Alivov, C. Liu, A. Teke, M.A. Reshchikov, S. Doğan, V. Avrutin, S.J. Cho, and H. Morkọ, J. Appl. Phys. **98**, 1 (2005). DOI:10.1063/1.1992666

[10] S.J. Pearton, *GaN and ZnO-Based Materials and Devices* (Springer Berlin Heidelberg, Berlin, Heidelberg, 2012).

[11] O. Madelung, *Semiconductors: Data Handbook* (Springer Berlin Heidelberg, Berlin, Heidelberg, 2004).

[12] K. Maschke and U. Rössler, *Phys. Status Solidi* **28**, 577 (1968). DOI:10.1002/pssb.19680280215.

[13] D.M. Roessler and W.C. Walker, *Phys. Rev*. **159**, 733 (1967). DOI:10.1103/PhysRev.159.733.

[14] G. Chen, K.M. Yu, L.A. Reichertz, and W. Walukiewicz, *Appl. Phys. Lett.* **103**, 1 (2013). DOI:10.1063/1.4816326.

[15] L.M. Guia, V. Sallet, C. Sartel, M.C. Martínez-Tomás, and V. Muñoz-Sanjosé, *Phys. Status Solidi Curr. Top. Solid State Phys.* **13**, 452 (2016). DOI:10.1002/pssc.201510276.

[16] L.M. Guia, V. Sallet, S. Hassani, M.C. Martínez-Tomás, and V. Muñoz-Sanjosé, *Cryst. Growth Des.* **17**, 6303 (2017). DOI: 10.1021/acs.cgd.7b00989.

[17] E. Przezdziecka, A. Wierzbicka, P. Dłuzewski, I. Sankowska, P. Sybilski, K. Morawiec, M.A. Pietrzyk, and A. Kozanecki, *Cryst. Growth Des.* **20**, 5466 (2020). DOI: 10.1021/acs.cgd.0c00678.

[18] E. Przeździecka, P. Strąk, A. Wierzbicka, A. Adhikari, A. Lysak, P. Sybilski, J.M. Sajkowski, A. Seweryn, and A. Kozanecki, *Nanoscale Res. Lett.* **16**, 59 (2021). DOI:10.1186/s11671-021-03517-y.

[19] A. Adhikari, A. Lysak, A. Wierzbicka, P. Sybilski, A. Reszka, B.S. Witkowski, and E. Przezdziecka, *Mater. Sci. Semicond. Process.* **144**, 106608 (2022). DOI: 10.1016/j.mssp.2022.106608.

[20] Y. Lee, C.P. Liu, K.M. Yu, and W. Walukiewicz, *Energy Technol.* **6**, 122 (2018). DOI:10.1002/ente.201700641.

[21] M. Stachowicz, M. Pietrzyk, D. Jarosz, P. Dluzewski, E. Alves, and A. Kozanecki, *Surf. Coatings Technol.* **355**, 45 (2018). DOI: 10.1016/j.surfcoat.2018.01.040.

[22] P. Bhattacharya, R.R. Das, and R.S. Katiyar, *Thin Solid Films* **447**–**448**, 564 (2004). DOI: 10.1016/j.tsf.2003.07.017.

[23] E. Przeździecka, A. Wierzbicka, A. Lysak, P. Dłużewski, A. Adhikari, P. Sybilski, K. Morawiec, and A. Kozanecki, *Cryst. Growth Des*. **22**, 1110 (2022). DOI: 10.1021/acs.cgd.1c01065.

[24] A. Lysak, E. Przeździecka, R. Jakiela, A. Reszka, B. Witkowski, Z. Khosravizadeh, A. Adhikari, J.M. Sajkowski, and A. Kozanecki, *Mater. Sci. Semicond. Process*. **142**, 1 (2022). DOI: 10.1016/j.mssp.2022.106493.

[25] K. Usharani, A.R. Balu, V.S. Nagarethinam, and M. Suganya, *Prog. Nat. Sci. Mater. Int.* **25**, 251 (2015). DOI : 10.1016/J.PNSC.2015.06.003.

[26] M.M. Sanjeevannanavar, J. Nettar, and D. D'Souza, *AIP Conf. Proc.* **2244**, 030002 (2020). DOI :10.1063/5.0009062.

[27] K. Bashir, N. Mehboob, M. Ashraf, A. Zaman, V. Tirth, A. Algahtani, A.U. Shah, A. Ali, M. Mushtaq, and K. Althubeiti, *J. Phys. Soc. Japan* **91**, (2022). DOI :10.7566/JPSJ.91.054706.

[28] B.A. Joyce, *Reports Prog. Phys.* **37**, 363 (1974). DOI :10.1088/0034-4885/37/3/002.

[29] J.M. Hinckley and J. Singh, *Phys. Rev. B* **42**, 3546 (1990). DOI:10.1103/PhysRevB.42.3546.

[30] P. Prepelita, R. Medianu, B. Sbarcea, F. Garoi, and M. Filipescu, *Appl. Surf. Sci.* **256**, 1807 (2010). DOI: 10.1016/J.APSUSC.2009.10.011.

[31] R.K. Gupta, K. Ghosh, R. Patel, and P.K. Kahol, *Appl. Surf. Sci.* **255**, 4466 (2009). DOI : 10.1016/J.APSUSC.2008.11.039.

[32] C. He, Q. Wu, X. Wang, Y. Zhang, L. Yang, N. Liu, Y. Zhao, Y. Lu, and Z. Hu, *ACS Nano* **5**, 1291 (2011). DOI:10.1021/nn1029845.

[33] J.I. Langford, *J. Appl. Crystallogr.* **11**, 10 (1978). DOI:10.1107/S0021889878012601.

[34] F.R.L. Schoening, *Acta Crystallogr*. **18**, 975 (1965). DOI:10.1107/S0365110X65002335.

[35] K. Usharani, A.R. Balu, V.S. Nagarethinam, and M. Suganya, *Prog. Nat. Sci. Mater. Int.* **25**, 251 (2015). DOI: 10.1016/j.pnsc.2015.06.003.

[36] A. Abdolahzadeh Ziabari, F.E. Ghodsi, and G. Kiriakidis, *Surf. Coatings Technol.* **213**, 15 (2012). DOI: 10.1016/J.SURFCOAT.2012.10.003.

[37] S. Karakaya and O. Ozbas, *AIP Conf. Proc.* **1569**, 253 (2013). DOI:10.1063/1.4849270.

[38] F. Atay, I. Akyuz, S. Kose, E. Ketenci, and V. Bilgin, *J. Mater. Sci. Mater. Electron.* **22**, 492 (2011). doi:10.1007/S10854-010-0166-Z.

[39] R.K. Gupta, F. Yakuphanoglu, and F.M. Amanullah, *Phys. E Low-Dimensional Syst. Nanostructures* **43**, 1666 (2011). DOI: 10.1016/j.physe.2011.05.019.

[40] A.S. Mohammed, D.K. Kafi, A. Ramizy, O.O. Abdulhadi, and S.F. Hasan, *J. Ovonic Res.* **15**, 37 (2019).

[41] F. Yakuphanoglu, *J. Sol-Gel Sci. Technol.* **59**, 569 (2011). DOI:10.1007/s10971-011-2528-2.

[42] T.G. Mayerhöfer, S. Pahlow, and J. Popp, *Chem Phys Chem* **21**, 2029 (2020). DOI:10.1002/CPHC.202000464.

[43] J. Tauc, Mater. Res. Bull. **3**, 37 (1968). DOI:10.1016/0025-5408(68)90023-8.

[44] K. Dileep, R. Sahu, S. Sarkar, S.C. Peter, and R. Datta, *J. Appl. Phys.* **119**, 114309 (2016). DOI:10.1063/1.4944431.

[45] S. Knief and W. von Niessen, *Phys. Rev. B* **59**, 12940 (1999). DOI:10.1103/PhysRevB.59.12940.

[46] K. Saito and A.J. Ikushima, *Phys. Rev. B* **62**, 8584 (2000). DOI:10.1103/PhysRevB.62.8584.

[47] R. Henríquez, P. Grez, E. Muñoz, H. Gómez, J.A. Badán, R.E. Marotti, and E.A. Dalchiele, *Thin Solid Films* **518**, 1774 (2010). DOI: 10.1016/J.TSF.2009.09.030.

[48] A.A. Dakhel, and F.Z. Henari, *Cryst. Res. Technol.* **38**, 979 (2003). DOI:10.1002/CRAT.200310124.

[49] K. Bashir, N. Mehboob, M. Ashraf, A. Zaman, F. Sultana, A.F. Khan, A. Ali, M. Mushtaq, S. Uddin, and K. Althubeiti, Materials (Basel). **14**, 4538 (2021). DOI:10.3390/ma14164538.

[50] S. Gowrishankar, L. Balakrishnan, and N. Gopalakrishnan, *Ceram. Int.* **40**, 2135 (2014). DOI: 10.1016/j.ceramint.2013.07.130.

[51] K. Usharani and A.R. Balu, *J. Mater. Sci. Mater. Electron.* **27**, 2071 (2016). DOI:10.1007/s10854-015-3993-0.

[52] K. Osamura, S. Naka, and Y. Murakami, *J. Appl. Phys.* **46**, 3432 (2008). DOI:10.1063/1.322064.

[53] A. Rajagopal, R.J. Stoddard, H.W. Hillhouse, and A.K.Y. Jen, *J. Mater. Chem. A* **7**, 16285 (2019). DOI:10.1039/C9TA05308E.

[54] A. Adhikari, E. Przezdziecka, S. Mishra, P. Sybilski, J. Sajkowski, and E. Guziewicz, *Phys. Status Solidi Appl. Mater. Sci.* **218**, 1 (2021). DOI:10.1002/pssa.202000669.

[55] J. Huso, H. Che, and D. Thapa, *J. Appl. Phys* **117**, 125702 (2015). DOI:10.1063/1.4916096.

[56] C. Kaiser, O.J. Sandberg, N. Zarrabi, W. Li, P. Meredith, and A. Armin, *Nat. Commun.* **12**, 1 (2021). DOI:10.1038/s41467-021-24202-9.

[57] V. Dalouji, and N. Rahimi, *Europhys. Lett.* **139**, 45002 (2022). DOI:10.1209/0295-5075/ac7890.

[58] R. Vettumperumal, S. Kalyanaraman, B. Santoshkumar, and R. Thangavel, *Mater. Res. Bull.* **77**, 101 (2016). DOI : 10.1016/j.materresbull.2016.01.015.

[59] R.C. Rai, *J. Appl. Phys*. **113**, 3508 (2013). DOI :10.1063/1.4801900.

[60] M.A. Islam, M.S. Hossain, M.M, Aliyu, P. Chelvanathan, Q. Huda, M.R. Karim, K. Sopian, and N. Amin, *Energy Procedia* **33**, 203 (2013). DOI: 10.1016/j.egypro.2013.05.059.

[61] S. Brahma, P.H. Lee, H.H. Chen, A.C. Lee, and J.L. Huang, *J. Phys. Chem. Solids* **148**, 109728 (2021). DOI: 10.1016/j.jpcs.2020.109728.

# Figures

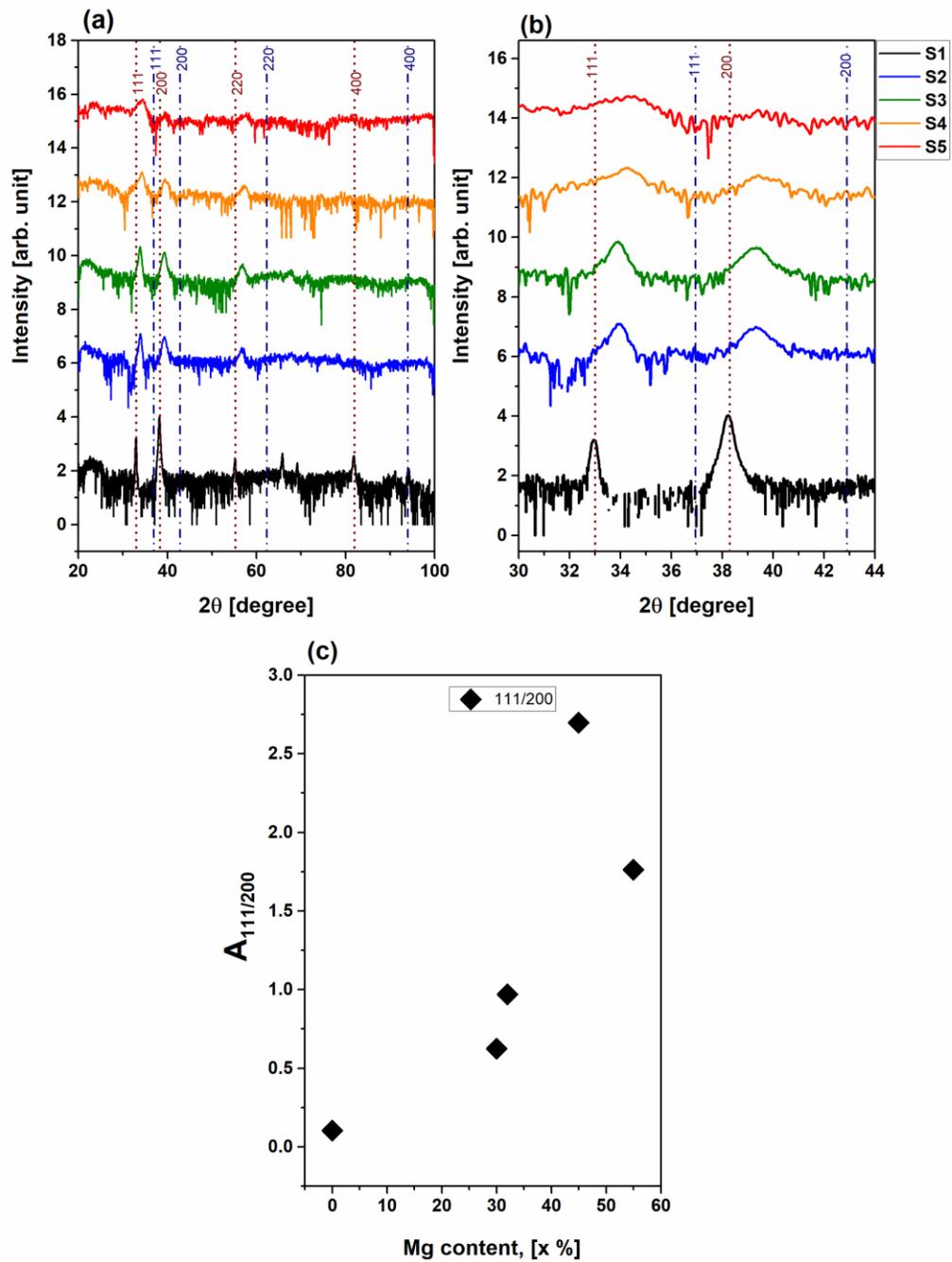

Fig. 1. (a) XRD pattern of CdMgO layers on a quartz substrate (b) 111 and 200 diffraction peaks shift (red and blue dotted line indicate different peak positions according to JCPDF card no. 652908 and JCPDF card no 870653 for CdO and MgO, respectively). (c) Ratio of the 111/200 peaks area.

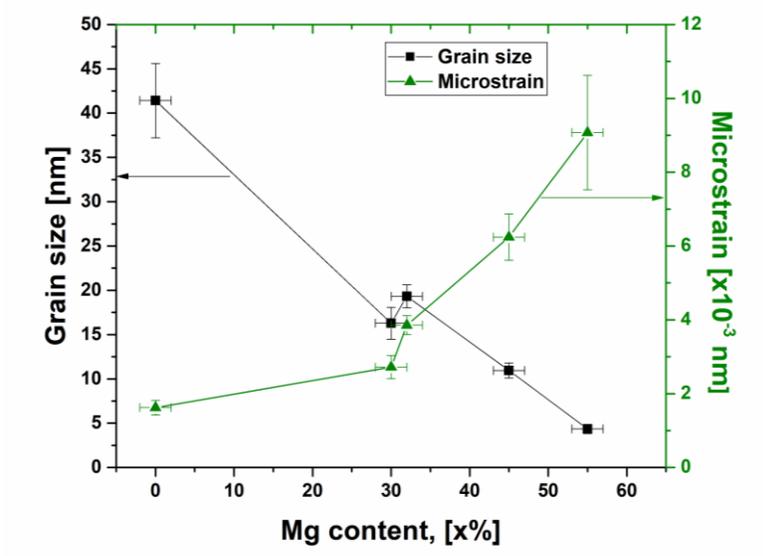

Fig. 2. Variation of Grain size and Microstrain in 111 diffraction peak direction with change in Mg content in CdMgO alloy.

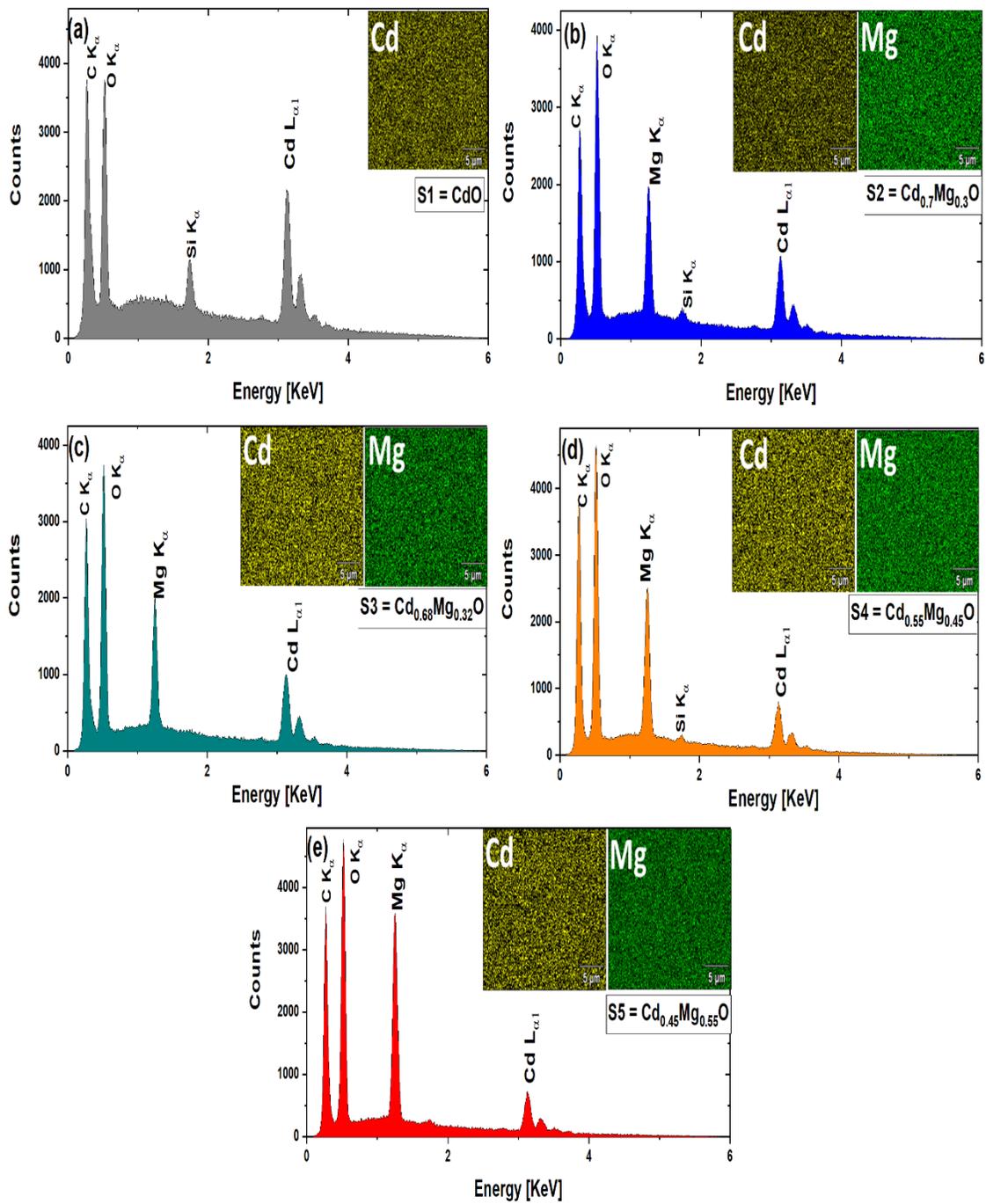

Fig. 3. EDX spectrum of (a) S1, (b) S2, (c) S3, (d) S4, and (e) S5 CdMgO alloys. The insert image shows the elemental distribution map of Cd and Mg in CdMgO alloys.

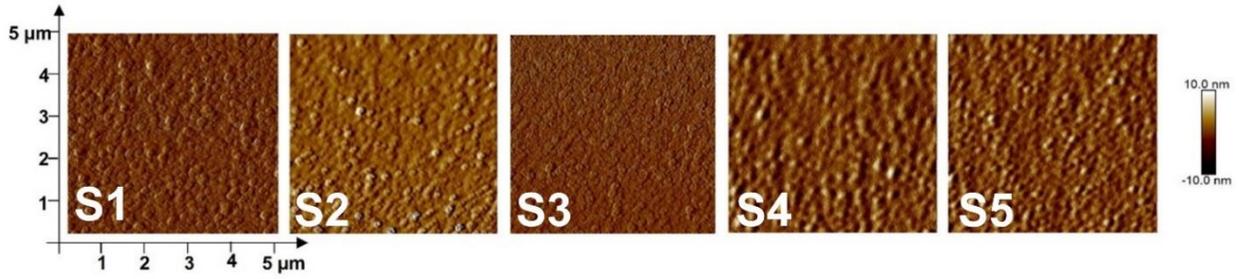

Fig. 4. AFM image of CdMgO layer with different Mg content.

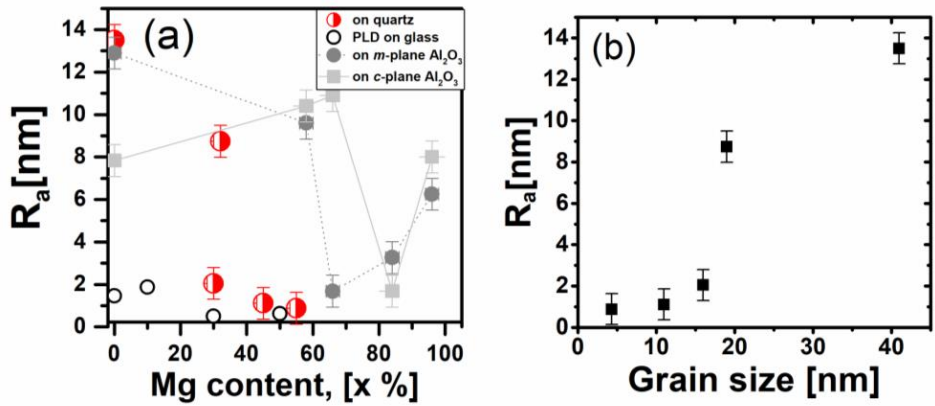

Fig. 5. (a) Roughness parameter vs Mg concentration (red point – this work, grey point-$R_a$ in case of samples grown on sapphire by MBE; based on [19], black points-on glass grown by PLD; based on [31]). (b) Variation of $R_a$ vs Grain size.

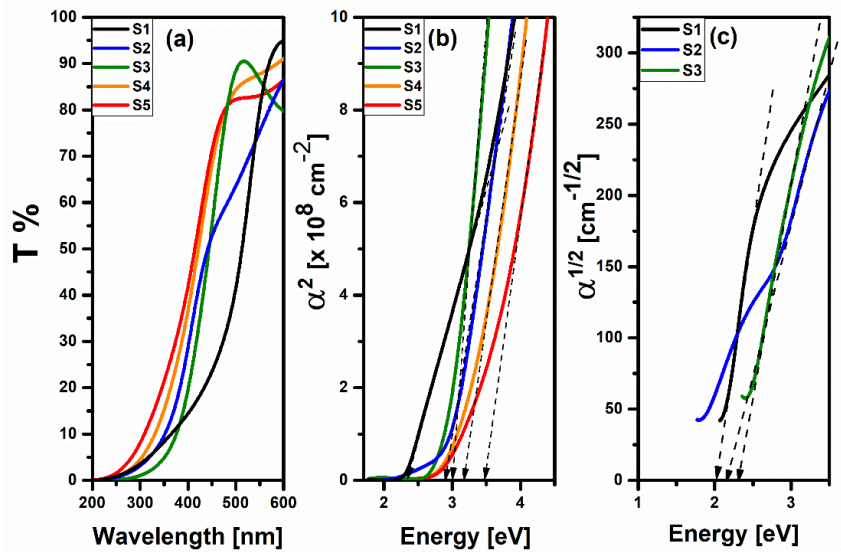

Fig. 6. (a) Transmittance spectra, (b) $\alpha^2$ plot as a function of energy (hv), and (c) $\alpha^{1/2}$ plot as a function of energy (hv).

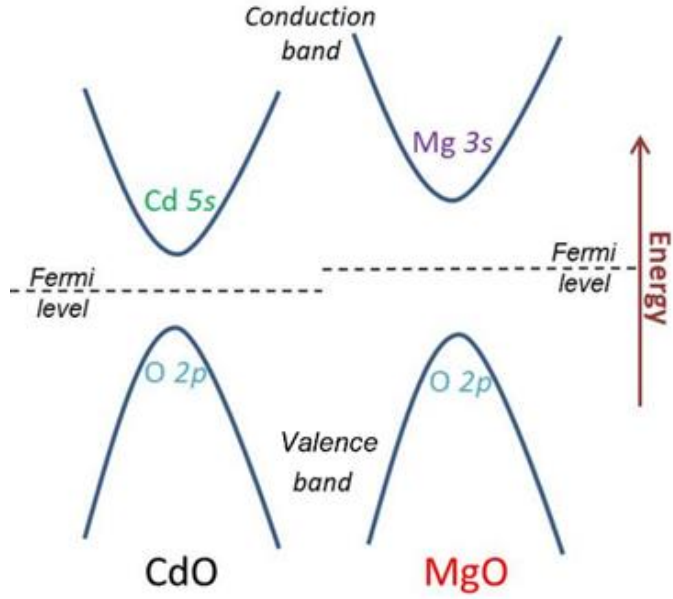

Fig. 7. Schematic illustration of the valence band and conduction band alignment in CdO and MgO.

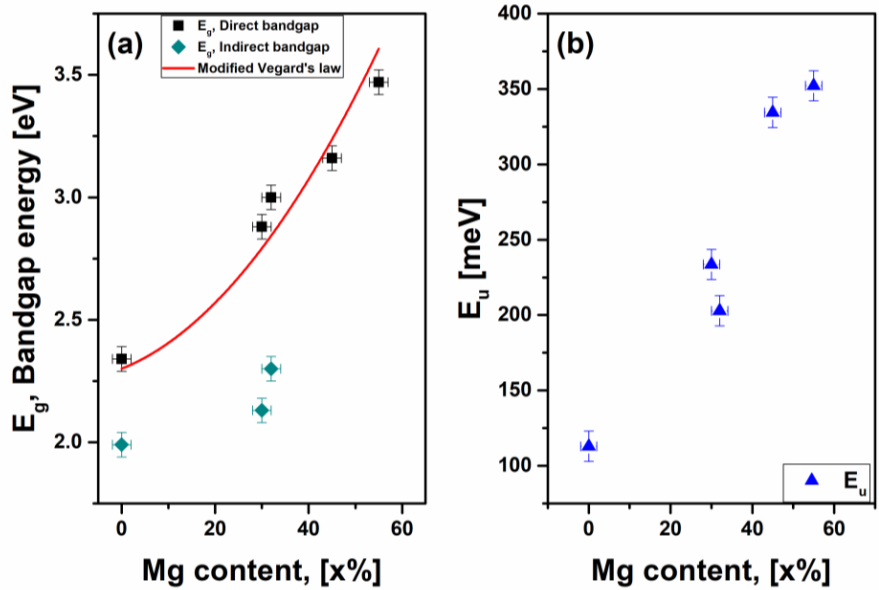

Fig. 8. Variation of (a) bandgap (black square and the cyan diamond represent direct and indirect bandgap respectively), and (b) Urbach Energy with change in Mg content

# Tables

**Table 1**. FWHM, lattice constant, grain size, and Microstrain of CdMgO alloys on quartz substrate along 111 diffraction peak.

| Sample name | 2θ [degrees] | FWHM [degrees] $\beta_G$ | FWHM [degrees] $\beta_L$ | Lattice parameter [nm] | Grain size [nm] | Microstrain, $\varepsilon$ [x$10^{-3}$nm] |
|---|---|---|---|---|---|---|
| S1 | 32.95 | 0.11 | 0.2 | 0.4703 | 41 | 1.6 |
| S2 | 33.92 | 0.19 | 0.51 | 0.4572 | 16 | 2.7 |
| S3 | 33.98 | 0.27 | 0.43 | 0.4564 | 19 | 3.9 |
| S4 | 34.21 | 0.44 | 0.76 | 0.4534 | 11 | 6.2 |
| S5 | 34.22 | 0.64 | 1.91 | 0.4533 | 4.3 | 9.07 |

**Table 2.** Cd and Mg content from EDX measurement, Mg effusion cell temperature, and average roughness parameter in the alloys.

| Sample name | Cd at% | Mg at% | Mg effusion cell temp [°C] | Average Roughness, $R_a$ [nm] |
|---|---|---|---|---|
| S1 | 100 | 0 | - | 13.5 |
| S2 | 70 | 30 | 500 | 2.05 |
| S3 | 68 | 32 | 510 | 8.74 |
| S4 | 55 | 45 | 520 | 1.11 |
| S5 | 45 | 55 | 530 | 0.88 |